\documentclass[a4paper]{article}

\pdfoutput=1

    \pdfcatalog {}

\usepackage{amsmath, amssymb, amsthm, mathrsfs, euscript}      
\usepackage{fancybox, fancyhdr, tocbibind}
\usepackage{ifthen}

    \usepackage[pdftex]{graphicx}
    \usepackage[usenames]{color}

    \definecolor{myDarkBlue}{rgb}{0, 0.04, 0.25}
    \definecolor{myDarkGreen}{rgb}{0, 0.3, 0}
    \definecolor{myDarkRed}{rgb}{0.3, 0, 0}

\makeatletter

\newcommand{\foot}[1]{%
  \edef\savecurrentlabel{\@currentlabel}%
                        \footnote{\hspace{1mm}%
                        \leftskip=-\parindent%
                        \ifthenelse{\value{footnote} < 10}%
			{\hangindent=6.7mm}{\hangindent=6.7mm}
                        #1}%
 \edef\@currentlabel{\savecurrentlabel}%
                       }
\newcommand{\pa}[1][]{\bigskip {\bf #1}}

\renewcommand{\paragraph}[1]{\pa\textbf{%
\mathversion{bold}#1\mathversion{normal}.}}

\newcommand{\expand}[2]{\begin{#1}#2\end{#1}}

\newcommand{\expandindex}[1]{\index{#1}}

\newcommand{\descriptions}[1]{{%
   \expand{description}{%
   \setlength{\leftskip}{7mm}%
   \smallskip #1\smallskip}%
}}

\newcommand{\form}[1] {\begin{equation}#1\end{equation}}
\newcommand{\formsplit}[1] {\form{\begin{split}#1\end{split}}}
\newcommand{\formskip}{\medskip}

\newcommand{\go}{\\[2.5mm]}
\newcommand{\eq}{\;=\;}
\newcommand{\fsequal}{&\eq}
\newcommand{\ap}{\;\approx\;}
\newcommand{\fsapprox}{&\ap}

\newcommand{\margincite}[1]{\marginpar[\centering #1]{\centering #1}}

\renewcommand\section{\thispagestyle{plain}%
                      \@startsection {section}{1}{\z@}%
                                   {-3.5ex \@plus -1ex \@minus -.2ex}%
                                   {2.3ex \@plus.2ex}%
                                   {\normalfont\Large\bfseries%
				   }}

\newcommand{\silentsection}[1]{
    \section*{#1}%
    \addcontentsline{toc}{section}{#1}%
    \@mkboth{\MakeUppercase{#1}}%
            {\MakeUppercase{#1}}%
    \thispagestyle{plain}\parindent\z@
}

\makeatother

\usepackage{makeidx}
\newcommand{\setS}{\ensuremath{{\mathcal{S}}}}

\newcommand{\kc}{Kolmogorov complexity}

\newcounter{ExperimentCounter}

\newcommand{\printtitle}{A Short Introduction to Kolmogorov Complexity}

\thicklines

\fancypagestyle{plain}{\fancyhead[C]{\printtitle}}

\makeatletter
\renewcommand\subsection{\@startsection {section}{1}{\z@}%
                                   {-3.5ex \@plus -1ex \@minus -.2ex}%
                                   {2.3ex \@plus.2ex}%
                                   {\normalfont\Large\bfseries%
				   }}
\renewcommand{\l@section}{\l@subsection}
\makeatother

\begin{document}
\thispagestyle{empty}
\parindent=0cm

\title{\printtitle\thanks{From {\em The Paradox of Overfitting}, \cite{Nannen:2003}.}}
\author{Volker Nannen}
\renewcommand{\today}{2003}

\maketitle

\begin{abstract}

\addcontentsline{toc}{section}{Abstract}

\noindent
This is a short introduction to \kc\ and information theory. 
The interested reader is referred to the
literature, especially the textbooks~\cite{Cover:1991}
and~\cite{Vitanyi:1997} which cover the fields of information theory
and \kc\ in depth and with all the necessary rigor. They are well to
read and require only a minimum of prior knowledge.

\end{abstract}
\thispagestyle{empty}

\paragraph{\kc}%
\index{Kolmogorov, A.N.}%
\index{Chaitin, G.J.}%
\index{Solomonoff, R.}%
\expandindex{\kc}%
\expandindex{algorithmic complexity|see{\kc}}%
\expandindex{Turing complexity|see{\kc}}%
\index{universal Turing machine}%
\index{UTM|see{universal Turing machine}} 
Also known as {\em algorithmic complexity\/} and {\em Turing
complexity}. Though Kolmogorov was not the first one to formulate the
idea, he played the dominant role in the consolidation of the theory.
The concept itself was developed independently and with different
motivation by Andrei N.~Kolmogorov \cite{Kolmogorov:1965}, Ray
Solomonoff \cite{Solomonoff:1964} and Gregory Chaitin
\cite{Chaitin:1966}, \cite{Chaitin:1969}. 

\pa

The \kc\ $C(s)$\margincite{$C(\cdot)$} of any binary string
$s\in\{0,1\}^n$ is the length of the shortest computer program $s^*$
that can produce this string on the Universal Turing Machine
UTM\margincite{UTM} and then halt. In other words, on the UTM $C(s)$
bits of information are needed to encode $s$.  The UTM is not a real
computer but an imaginary reference machine. We don't need the
specific details of the UTM. As every Turing machine can be
implemented on every other one, the minimum length of a program on one
machine will only add a constant to the minimum length of the program
on every other machine. This constant is the length of the
implementation of the first machine on the other machine and is
independent of the string in question.  This was first observed in
1964 by Ray Solomonoff.

\pa

Experience has shown that every attempt to construct a theoretical
model of computation that is more powerful than the Turing machine has
come up with something that is at the most just as strong as the
Turing machine. This has been codified in 1936 by Alonzo Church as
\index{Church's Thesis} Church's Thesis: the class of algorithmically 
computable numerical functions coincides with the class of partial
recursive functions.  Everything we can compute we can compute by a
Turing machine and what we cannot compute by a Turing machine we
cannot compute at all. This said, we can use \kc\ as a universal
measure that will assign the same value to any sequence of bits
regardless of the model of computation, within the bounds of an
additive constant.

\paragraph{Incomputability of \kc}
\kc\ is not computable. It is nevertheless essential for proving
existence and bounds for weaker notions of complexity. The fact that
\kc\ cannot be computed stems from the fact that we cannot compute
the output of every program. More fundamentally, no algorithm is
possible that can predict of every program if it will ever halt, as
has been shown by Alan Turing in his famous work on the halting
problem~\cite{Turing:1936}. No computer program is possible that, when
given any other computer program as input, will always output
\verb_true_ if that program will eventually halt and \verb_false_ if
it will not. Even if we have a short program that outputs our string
and that seems to be a good candidate for being the shortest such
program, there is always a number of shorter programs of which we do
not know if they will ever halt and with what output.

\paragraph{Plain versus prefix complexity}
Turing's original model of computation included special delimiters that
marked the end of an input string. This has resulted in two brands of
\kc: 

\descriptions{

  \item[plain \kc:]\expandindex{\kc!plain}\index{C@$C(\cdot)$}
      the\margincite{$C(\cdot)$} length $C(s)$ of the shortest binary
      string that is delimited by special marks and that can compute
      $x$ on the UTM and then halt.

  \item[prefix \kc:]\expandindex{\kc!prefix}\index{K@$K(\cdot)$}
       the\margincite{$K(\cdot)$} length $K(s)$ of the shortest binary
       string that is {\em self-delimiting\/}~\cite{Vitanyi:1997} and
       that can compute $x$ on the UTM and then halt.

}

The difference between the two is logarithmic in $C(s)$: the number of
extra bits that are needed to delimit the input string. While plain
\kc\ integrates neatly with the Turing model of computation,
prefix \kc\ has a number of desirable mathematical
characteristics that make it a more coherent theory. The individual
advantages and disadvantages are described in~\cite{Vitanyi:1997}.
Which one is actually used is a matter of convenience. We will mostly
use the prefix complexity $K(s)$.

\paragraph{Individual randomness}%
\index{individual randomness}
A. N. Kolmogorov was interested in \kc\ to define the individual
randomness of an object.  When $s$ has no computable regularity it
cannot be encoded by a program shorter than $s$. Such a string is
truly random and its \kc\ is the length of the string itself plus the
commando {\sf print}\foot{Plus a logarithmic term if we use prefix
complexity}. And indeed, strings with a \kc\ close to their actual
length satisfy all known tests of randomness. A regular string, on the
other hand, can be computed by a program much shorter than the string
itself.  But the overwhelming majority of all strings of any length
are random and for a string picked at random chances are exponentially
small that its \kc\ will be significantly smaller than its actual
length.

\pa

This can easily be shown. For any given integer $n$ there are exactly
$2^n$ binary strings of that length and $2^n-1$ strings that are
shorter than $n$: one empty string, $2^1$ strings of length one, $2^2$
of length two and so forth. Even if all strings shorter than $n$ would
produce a string of length $n$ on the UTM we would still be one string
short of assigning a $C(s)<n$ to every single one of our $2^n$
strings. And if we want to assign a $C(s)<n-1$ we can maximally do so
for $2^{n-1}-1$ strings. And for $C(s)<n-10$ we can only do so for
$2^{n-10}-1$ strings which is less than $0.1\%$ of all our strings.  Even
under optimal circumstances we will never find a $C(s)<n-c$ for more
than $\frac{1}{2^c}$ of our strings.

\paragraph{Conditional \kc}%
\expandindex{\kc!conditional}%
\index{K@$K(\cdot"|\cdot)$}
The conditional \kc\ $K(s|a)$\margincite{$K(\cdot|\cdot)$} is defined
as the shortest program that can output $s$ on the UTM if the input
string $a$ is given on an auxiliary tape. $K(s)$ is the special case
$K(s|\epsilon)$ where the auxiliary tape is empty. 

%It deserves special attention that
%
%\form{
%  K\big(\,s\,|\,K(s)\,\big)\; =\; K(s) - K\big(\,K(s)\,\big)
%}
%
%\pa
%
%In other words, the information that the prefix complexity $K(s)$
%contains on $s$ is about the extra number of bits that we need to
%delimit the shortest program $s^*$, and nothing more.

\paragraph{The universal distribution}%
\index{universal distribution}%
\index{Bayes' rule}%
\index{Solomonoff, R.}
When Ray Solomonoff first developed \kc\ in 1964 he intended it to
define a universal distribution over all possible objects. His
original approach dealt with a specific problem of Bayes' rule, the
unknown prior distribution. \index{distribution!prior}
\index{prior|see{distribution}} Bayes' rule can be used to calculate
$P(m|s)$, the probability for a probabilistic model to have generated
the sample $s$, given $s$. It is very simple. $P(s|m)$, the
probability that the sample will occur given the model, is multiplied
by the unconditional probability that the model will apply at all,
$P(m)$. This is divided by the unconditional probability of the sample
$P(s)$. The unconditional probability of the model is called the prior
distribution and the probability that the model will have generated
the data is called the posterior distribution.
\index{posterior|see{distribution}} \index{distribution!posterior}

\form{
  P(m|s) \eq \frac{P(s|m)\;P(m)}{P(s)}
}

\pa

Bayes' rule can easily be derived from the definition of conditional
probability:

\form{
  P(m|s) \eq \frac{P(m,s)}{P(s)}
}
and
\form{
  P(s|m) \eq \frac{P(m,s)}{P(m)}
}

\pa

The big and obvious problem with Bayes' rule is that we usually have
no idea what the prior distribution $P(m)$ should be. Solomonoff suggested
that if the true prior distribution is unknown the best assumption
would be the universal distribution $2^{-K(m)}$ where $K(m)$ is the
\expandindex{\kc!prefix} prefix \kc\ of the
model\foot{\expandindex{\kc!plain}Originally Solomonoff used the plain
\kc\ $C(\cdot)$. This resulted in an improper distribution $2^{-C(m)}$
that tends to infinity. Only in 1974 L.A. Levin introduced prefix
complexity to solve this particular problem, and thereby many other
problems as well~\cite{Levin:1974}.}.  This is nothing but a modern
codification of the age old principle that is wildly known under the
name of \index{Occam's razor} Occam's razor: the simplest explanation
is the most likely one to be true.

\paragraph{Entropy}%
\index{Shannon, C.E.}%
\index{entropy}%
\index{code length}
Claude Shannon~\cite{Shannon:1948} developed information
theory\index{information theory} in the late 1940's. He was concerned
with the optimum code length that could be given to different binary
words $w$ of a source string $s$. Obviously, assigning a short code
length to low frequency words or a long code length to high frequency
words is a waste of resources. Suppose we draw a word $w$ from our
source string $s$ uniformly at random. Then the probability $p(w)$ is
equal to the frequency of $w$ in $s$.  Shannon found that the optimum
overall code length for $s$ was achieved when assigning to each word
$w$ a code of length $-\log p(w)$. Shannon attributed the original
idea to R.M. Fano and hence this code is called the
Shannon-Fano\index{Shannon-Fano code} code.  When using such an
optimal code, the average code length of the words of $s$ can be
reduced to

\form{
  H(s) \eq -\sum_{w\in s} p(w)\log p(w)
}

\formskip

where $H(s)$\margincite{$H(\cdot)$}\index{H@$H(\cdot)$} is called the
entropy of the set $s$. When $s$ is finite and we assign a code of
length $-\log p(w)$ to each of the $n$ words of $s$, the total code
length is

\form{
  -\sum_{w\in s}\log p(w) \eq n\, H(s)
}

\pa

Let $s$ be the outcome of some random process $W$ that produces the
words $w\in s$ sequentially and independently, each with some known
probability \mbox{$p(W=w)>0$}.  $K(s|W)$ is the \kc\ of $s$ given $W$.
Because the Shannon-Fano code is optimal, the probability that
$K(s|W)$ is significantly less than $n H(W)$ is exponentially
small. This makes the negative log likelihood of $s$ given $W$ a good
estimator of $K(s|W)$:

\formsplit{
  K(s|W) 
\fsapprox
  n\,H(W)
\go\fsapprox
  \sum_{w\in s}\log p(w|W)
\go\fsequal
  -\log p(s|W)
}

\paragraph{Relative entropy}%
\label{relative-entropy}%
\index{entropy!relative|see{Kullback Leibler distance}}%
\index{Kullback Leibler distance}%
\index{D@$D(\cdot"|"|\cdot)$}
The relative entropy $D(p||q)$\margincite{$D(\cdot||\cdot)$} tells us
what happens when we use the wrong probability to encode our source
string $s$.  If $p(w)$ is the true distribution over the words of $s$
but we use $q(w)$ to encode them, we end up with an average of
$H(p)+D(p||q)$ bits per word. $D(p||q)$ is also called the Kullback
Leibler distance between the two probability mass functions $p$ and
$q$. It is defined as

\form{
  D(p||q) \eq \sum_{w\in s}p(w)\log\frac{p(w)}{q(w)}
}

\paragraph{Fisher information}%
\label{fisher}%
\index{Fisher information}
Fisher information was introduced into statistics some 20 years before
C. Shannon introduced information theory~\cite{Fisher:1925}. But it
was not well understood without it. Fisher information is the variance
of the score $V$ of the continuous parameter space of our models
$m_k$. This needs some explanation. At the beginning of this thesis we
defined models as binary strings that discretize the parameter space
of some function or probability distribution. For the purpose of
Fisher information we have to temporarily treat a model $m_k$ as a
vector in $\mathbb{R}^k$. And we only consider models where for all
samples $s$ the mapping $f_s(m_k)$ defined by $f_s(m_k)=p(s|m_k)$ is
differentiable. Then the score $V$ can be defined as

\formsplit{
  V 
\fsequal
  \frac{\partial}{\partial\, m_k}\; \ln p(s|m_k)
\go\fsequal
  \frac{\frac{\partial}{\partial\, m_k}\;p(s|m_k)}{p(s|m_k)}
}

\pa

The score $V$ is the partial derivative of $\ln\,p(s|m_k)$, a term we
are already familiar with. The Fisher information
$J(m_k)$\index{J@$J(\cdot)$}\margincite{$J(\cdot)$} is

\form{ 
  J(m_k) \eq E_{m_k}\left[\frac{\partial}{\partial\,m_k}\;\ln
  p(s|m_k) \right]^2 
}

\pa

Intuitively, a high Fisher information means that slight changes to
the parameters will have a great effect on $p(s|m_k)$.  If $J(m_k)$ is
high we must calculate $p(s|m_k)$ to a high
precision\index{precision}\index{rounding}. Conversely, if $J(m_k)$ is
low, we may round $p(s|m_k)$ to a low precision.

\paragraph{\kc\ of sets}%
\expandindex{\kc!of sets}
The \kc\ of a set of strings \setS\ is the length of the shortest
program that can output the members of \setS\ on the UTM and then
halt.  If one is to approximate some string~$s$ with $\alpha < K(s)$
bits then the best one can do is to compute the smallest set $\setS$
with $K(\setS)\le\alpha$ that includes $s$. Once we have some
$\setS\ni s$ we need at most $\log|\setS|$ additional bits to
compute~$s$. This set \setS\ is defined by the Kolmogorov structure
function\index{Kolmogorov structure
function}\margincite{$h_s(\cdot)$}\index{h@$h_s(\cdot)$}

\form{
  h_s(\alpha) \eq \operatornamewithlimits{min}_\setS\big[\log
     |\setS| : \setS\ni s,\; K(\setS)\le\alpha\big]
}

\formskip

which has many interesting features. The function $h_s(\alpha)+\alpha$
is non increasing and never falls below the line $K(s)+O(1)$ but can
assume any form within these constraints. It should be evident that

\form{\label{ksf-set}
  h_s(\alpha) \;\ge\; K(s) - K(\setS)
}

\paragraph{\kc\ of distributions}%
\expandindex{\kc!of distributions}
The Kolmogorov structure function is not confined to finite sets.  If
we generalize $h_s(\alpha)$ to probabilistic models $m_p$ that define
distributions over $\mathbb R$ and if we let~$s$ describe a real
number, we obtain

\form{\label{ksf-distribution}
  h_s(\alpha) \eq \operatornamewithlimits{min}_{m_p}\big[-\log p(s|m_p)
     : p(s|m_p)>0,\; K(m_p)\le\alpha\big]
}

\formskip

where $-\log p(s|m_p)$ is the number of bits we need to encode~$s$
with a code that is optimal for the distribution defined
by~$m_p$. Henceforth we will write $m_p$
%\margincite{$m_p$} 
when the model\index{model} defines a probability distribution and
$m_k$
%\margincite{$m_k$}
with $k\in\mathbb{N}$ when the model defines a probability
distribution that has $k$ parameters.  A set \setS\ can be viewed
as a special case of $m_p$, a uniform distribution with

\form{
 p(s|m_p) \eq \begin{cases}
   \; \frac{1}{|\setS|}   &   \text{if }\; s\in\setS 
\go
   \; 0              &   \text{if }\; s\not\in\setS
 \end{cases}
}

\paragraph{Minimum randomness deficiency}%
\label{min-ran-def}%
\index{minimum randomness deficiency}%
\index{randomness deficiency}
The randomness deficiency of a string~$s$ with regard to a model $m_p$
is defined as\margincite{$\delta(\cdot|m_p)$}\index{d@$\delta(\cdot"|m_p)$}

\form{
  \delta(s|m_p) \eq -\log p(s|m_p) - K(\,s|m_p,\, K(m_p)\,)
}

\formskip

for $p(s)>0$, and $\infty$ otherwise. This is a generalization of the
definition given in~\cite{Vitanyi:2002} where models are finite
sets. If $\delta(s|m_p)$ is small, then~$s$ may be considered a {\em
typical\/} or {\em low profile\/} instance of the distribution. $s$
satisfies {\em all\/} properties of low \kc\ that hold with high
probability for the support set of~$m_p$. This would not be the case
if $s$ would be exactly identical to the mean, first momentum or any
other special characteristic of $m_p$.

\pa

Randomness deficiency is a key concept to any application of \kc. As
we saw earlier, \kc\ and conditional \kc\ are not computable. We can
never claim that a particular string $s$ does have a conditional \kc\

\form{
  K(s|m_p) \ap -\log p(s|m_p)
}

\formskip

The technical term that defines all those strings that do satisfy this
approximation is {\em
typicality\index{typicality}\margincite{typicality}}, defined as a
small randomness deficiency $\delta(s|m_p)$.

\pa

Minimum randomness deficiency turns out to be important for lossy data
compression.  A compressed string of minimum randomness deficiency is
the most difficult one to distinguish from the original string. The
best lossy compression that uses a maximum of $\alpha$ bits is defined
by the minimum randomness deficiency
function

\margincite{$\beta_s(\cdot)$}\index{b@$\beta_s(\cdot)$}

\form{
  \beta_s(\alpha) \eq \operatornamewithlimits{min}_{m_p}\big[\delta(s|m_p):
  p(s|m_p)>0,\; K(m_p)\le\alpha\big] }

\paragraph{Minimum Description Length}%
\index{MDL}%
\index{minimum description length|see{MDL}}
The Minimum Description Length or short MDL\margincite{MDL} of a
string~$s$ is the length of the shortest two-part code for~$s$ that
uses less than $\alpha$ bits. It consists of the number of bits needed
to encode the model $m_p$ that defines a distribution and the negative
log likelihood of~$s$ under this
distribution.\margincite{$\lambda_s(\cdot)$}\index{l@$\lambda_s(\cdot)$}

\form{
  \lambda_s(\alpha) \eq \operatornamewithlimits{min}_{m_p}\big[
             -\log p(s|m_p) + K(m_p):
             p (s|m_p)>0,\;K(m_p)\le\alpha]
}

\pa

It has recently been shown by Nikolai Vereshchagin and Paul Vit\'anyi
in~\cite{Vitanyi:2002} that a model that minimizes the description
length also minimizes the randomness deficiency, though the reverse
may not be true. The most fundamental result of that paper is the
equality

\form{
    \beta_s(\alpha) 
  \eq 
    h_s(\alpha)+\alpha-K(s) 
  \eq
    \lambda_s(\alpha)-K(s)
}

\formskip

where the mutual relations between the Kolmogorov structure function,
the minimum randomness deficiency and the minimum description length
are pinned down, up to logarithmic additive terms in argument and
value.

\bibliographystyle{alpha}

\end{document}